\documentclass[prl,floatfix,twocolumn]{revtex4-1}
\usepackage{amsmath,amsfonts,amssymb}
\usepackage{mathrsfs}
\usepackage{graphicx}
\usepackage{epsfig}
\usepackage{epstopdf}
\usepackage{hyperref}
\usepackage{xcolor}
\usepackage{ulem}
\usepackage{placeins}
\usepackage{graphicx}
\usepackage{dcolumn}
\usepackage{gensymb}

\begin{document}

\title{The nonlinearity of interactions drives networks of neural oscillators to decoherence at strong coupling}

\author{Richa Tripathi$^{1}$, Shakti N. Menon$^2$ and Sitabhra Sinha$^{2,3}$}

\affiliation{$^1$ Indian Institute of Technology Gandhinagar, Gandhinagar, Gujarat 382355, India.\\
$^2$The Institute of Mathematical Sciences, CIT Campus,
Taramani, Chennai 600113, India.\\
$^3$Homi Bhabha National Institute, Anushaktinagar, Mumbai 400094, India.}
\date{\today}
 
\begin{abstract}
While phase oscillators are often used to model neuronal populations, in contrast to the Kuramoto paradigm, strong interactions between brain areas can be associated with loss of synchrony. Using networks of coupled oscillators described by neural mass models, we find that a transition to decoherence at increased coupling strength results from the fundamental nonlinearity, e.g., arising from refractoriness, of the interactions between the nodes. The nonlinearity-driven transition also depends on the connection topology, underlining the role of network structure in shaping brain activity.
\end{abstract}

\maketitle

Collective oscillations in
large populations of synaptically coupled neurons provide a striking example of
the rich diversity of complex behavior that can arise through nonlinear interactions in the brain~\cite{Buzsaki2004,Buzsaki2006}. 
These emergent phenomena are known to have functional consequences~\cite{Lakatos2008}, as in the case of coherent activity achieved via neural synchronization~\cite{Rodriguez1999,Engel2001,Fries2001}.
One of the simplest models used to investigate the transition to coherence in systems of interacting oscillators
is the one proposed by Kuramoto~\cite{Kuramoto1984,Acebron2005,Rodrigues2016},
which provides a natural framework for describing recurrent activity in the brain~\cite{Rabinovich2006}. 
The model shows that a population of heterogeneous oscillators that are globally coupled with sufficient strength can achieve coherence.
The robustness of this transition across different types of heterogeneity has enabled the model to be used
for describing synchronization
in a large variety of natural systems~\cite{Pikovsky2003} 
including the brain~\cite{Montbrio2018} 
However, there are aspects of neuronal collective dynamics that do not appear to be in accordance with the paradigm of global 
synchronization arising at large coupling strengths. 

Studies have shown that
while loss of consciousness is associated with increased synchronization among brain areas~\cite{Kar2011,Chu2012,Li2013,Bola2018}, 
the interaction between them concurrently decreases~\cite{Lewis2012,Schroeder2016}, suggesting that \textit{increased coupling is accompanied by decreased coherence} in the brain.
This phenomenon, which runs counter to the transition to synchrony expected from the Kuramoto paradigm,
raises a basic question: can such contrary behavior be associated with
the presence of fundamentally nonlinear interactions in the brain? 
{\it Neural mass models} provide a natural framework for investigating the dynamical consequences of nonlinear coupling between
brain regions. In such models, the 
activity of a large number of
neurons interacting via synapses is reduced to an aggregate description of the
dynamics of specific subpopulations~\cite{DaSilva1974,Freeman1978,Jansen1995,Deco2008}.
The Wilson-Cowan (WC) model, perhaps the best known model of this type,
describes the activity at a local region 
of the cortex in terms of interactions between two distinct subpopulations comprising 
excitatory and inhibitory neurons respectively~\cite{Wilson1972,Destexhe2009}.
The derivation of a phase description of the WC model under extremely restrictive assumptions
has been used to assert that its
synchronization behavior is equivalent to that
of the Kuramoto class of models~\cite{Schuster1990a,Schuster1990b,Hoppensteadt1997,Daffertshofer2011}.
However, this correspondence between the two models breaks down under biologically realistic
conditions, in particular, when we 
explicitly consider refractoriness, i.e., the insensitivity
of neurons to stimuli for a finite duration following excitation,
which makes the interactions between the WC oscillators fundamentally nonlinear even for weak coupling.
In this paper we show that this nonlinearity causes the dynamics of networks of neural mass models to diverge radically from the Kuramoto paradigm
with, most importantly, stronger interactions between the nodes promoting decoherence, consistent with
observations in the brain.
An additional consequence of the intrinsic nonlinearity of the system is that, unlike coupled phase oscillators, the effect
of the connection topology on the collective behavior is distinctly manifested.   
The difference with the Kuramoto model is further underlined by our observation that the emergent frequency of the coupled WC oscillator system increases monotonically with the coupling strength, even beyond the individual intrinsic frequencies of the nodes. 
Our results suggest a deeper appreciation of nonlinear interactions in complex systems that can invert the dynamical
behavior expected from systems with linearizable couplings.


%
%

\begin{figure}
\includegraphics{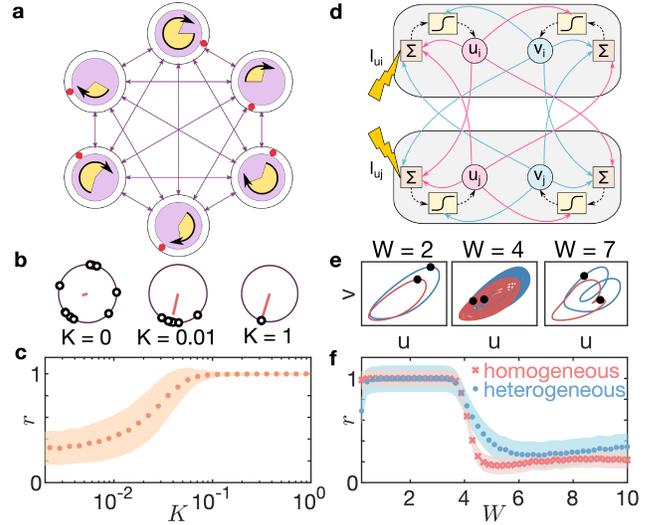}
\caption{{\bf Increasing the interaction strength leads to diverging outcomes in the emergent
collective dynamics of coupled phase oscillators and that of nonlinearly interacting neural
oscillators.}
(a)~A system of globally coupled phase oscillators (described by the Kuramoto model) shown schematically 
where the state of each element (disc) is characterized by its instantaneous
phase (position of the bead on the surrounding ring). Heterogeneity among the elements 
is indicated by the lengths of the directed arcs, which represent
the intrinsic oscillator frequency $\omega$, being different across discs.
(b-c) Emergence of coherence with increasing coupling strength $K$ in a system
of $N=10$ oscillators is (b) indicated using snapshots of the instantaneous state at specific values of
$K$, with the phases of individual elements indicated as beads on the ring, and 
(c) quantified using the phase coherence order parameter $r$. The value of
$r$ for each case in (b) is schematically represented 
by the lengths of the line segments inside the rings and $\omega$
are sampled from a Cauchy~($0,  0.0006$) distribution. 
(d) Schematic representation of a pair of nonlinearly coupled Wilson-Cowan (WC) oscillators
(shaded boxes), each comprising subpopulations of excitatory ($u$) and
inhibitory ($v$) cells. The state of each compartment $u_i, v_i$ in an oscillator is a sigmoid function 
of the weighted sum 
of the inputs received from other compartments to which it is coupled. The lightning
bolt symbols represent the external stimuli $I_u$ applied to the excitatory
subpopulation of each oscillator.
(e) Collective dynamics of two coupled WC oscillators represented by the phase-plane trajectories of the
oscillators which have different intrinsic 
frequencies
($\omega_1 = 0.975$, $\omega_2 = 1.031$, expressed relative to the reference frequency
$\omega_0$ [see text]). On increasing the coupling strength $W$ between the oscillators, we observe 
[L-R] inhomogeneous in-phase synchronization (IIS), quasiperiodic activity (QP)
and inhomogeneous anti-phase synchronization (IAPS). The instantaneous positions
of the two oscillators in phase space is indicated by the beads, and $I_u = 1.25$ in all cases.
(f) Strong coupling leads to loss of coherence, as indicated by the decrease in the order parameter $r$ at large $W$ 
in a system of $N=10$ globally coupled WC oscillators, irrespective of whether they have the same $\omega$ (homogeneous)
or are heterogeneous.
In panels (c) and (f), the filled circles and shaded regions represent the
means and standard deviations computed over $1000$ and $400$ realizations, respectively.
}
\label{fig:fig1}
\end{figure}

In Fig.~\ref{fig:fig1}, we contrast the behavior of the Kuramoto model of coupled phase oscillators (a-c) with that of the nonlinearly
coupled WC neural mass model (d-f).
Fig.~\ref{fig:fig1}~(a) schematically represents a globally coupled system of $N$
oscillators, whose instantaneous 
state is specified only by their phase. Heterogeneity among the units is introduced by choosing the intrinsic frequencies $\omega_{j}$ ($j=1,\ldots, N$) 
from a distribution (typically a Lorentzian). 
The time-evolution of the phases $\varphi_{j}$ of the oscillators are described using the Kuramoto model, viz.,
$\dot \varphi_{j} = \omega_{j} + (K/N)\Sigma^N_{i=1} \sin(\varphi_{i}-\varphi_{j})$. As the
coupling strength $K$ is increased, the difference between the phases reduces until
the system reaches a state of exact synchronization (ES) of all the oscillators (as shown in
Fig.~\ref{fig:fig1}~(b)). This transition is illustrated in Fig.~\ref{fig:fig1}~(c) in
terms of the coherence order parameter $r = N^{-1} \left |\sum_{j = 1}^{N}\exp^{i\varphi_{j}}\right |$, where $r=1$ corresponds
to ES.
 
Qualitatively distinct behavior is shown by systems comprising $N$ WC oscillators coupled to each other, 
as shown schematically for a pair of nodes in Fig.~\ref{fig:fig1}~(d).
Each unit $i$ represents a brain region
whose dynamical state is specified by the variables $u_i(t)$ and $v_i(t)$, characterizing the aggregate activity in the interacting 
subpopulations of excitatory and inhibitory neurons, respectively, and which evolve as:
\begin{equation} \label{wc_eqn}
\begin{split}
\tau_u \dot{u}_i &= -u_i + (\kappa_u - r_u u_i)\ {\cal S}_u(u_i^{in}), \\
\tau_v \dot{v}_i &= -v_i + (\kappa_v - r_v v_i)\ {\cal S}_v(v_i^{in}).
\end{split}
\end{equation}
The sigmoidal function $\mathcal{S}_{\mu}(x) = \kappa_{\mu} - 1+[1+ \exp\{-a_{\mu}(x-\theta_{\mu})\}]^{-1}$, which saturates to a maximum
value of $\kappa_{\mu} = 1- [1+ \exp({a_{\mu}\theta_{\mu}})]^{-1}$, describes
the response of each of the subpopulations to their net stimulation (input), viz., $u_{i}^{in} = c_{uu}u_{i} - c_{uv}v_{i}+ \Sigma_j (W_{ij}^{uu}u_{j} - W_{ij}^{uv}v_{j})  +I_{u}$ and
$v_{i}^{in} = c_{vu}u_{i} - c_{vv}v_{i}+ \Sigma_j (W_{ij}^{vu}u_{j} - W_{ij}^{vv}v_{j}) +I_{v}$.
The strength of interactions within and between the $u,v$ subpopulations of each node are parametrized by the constants $c_{\mu \nu} (\mu,\nu=u,v)$, 
while the time-invariant external stimuli received by them are represented 
by $I_{u} (=1.25)$ and $I_{v} (=0)$, respectively. 
The different subpopulations of each pair $i,j$ of connected nodes are assumed to be coupled with the same strength, viz.,  
$W_{ij}^{\mu\nu}= W/(N-1)$ $(\mu,\nu=u,v)$  $\forall i\neq j$ ($W_{ii}=0$). 
Hence, in our model the coupling between oscillators is specified by a single parameter, $W$.
Parameter values are chosen such that each node exhibits autonomous oscillations, viz., $a_{u}=1.3, \theta_{u}=4, a_{v}=2, 
\theta_{v}=3.7, 
r_{u}=1, r_{v}=1$,  $\tau_{u}=\tau_{v} = 8$.
We would like to note that $r_u, r_v$ correspond to refractory periods of the neurons and the choice of a finite value
makes the interactions irreducibly nonlinear (see SI).
For homogeneous systems of oscillators we have chosen $c_{uu}=16, c_{uv}=12, c_{vu}=15, c_{vv}=3$, which corresponds  to each
node in isolation oscillating with intrinsic frequency $\omega_0$ ($= 0.025$ arb. units, considered as the reference value relative to which all other
frequencies are expressed). For heterogeneous systems, each node has a different intrinsic frequency which is a consequence of randomly
sampling the values of $c_{\mu \nu}$ for each oscillator from log-normal distributions (see SI).

%
%
%

\begin{figure}
\includegraphics{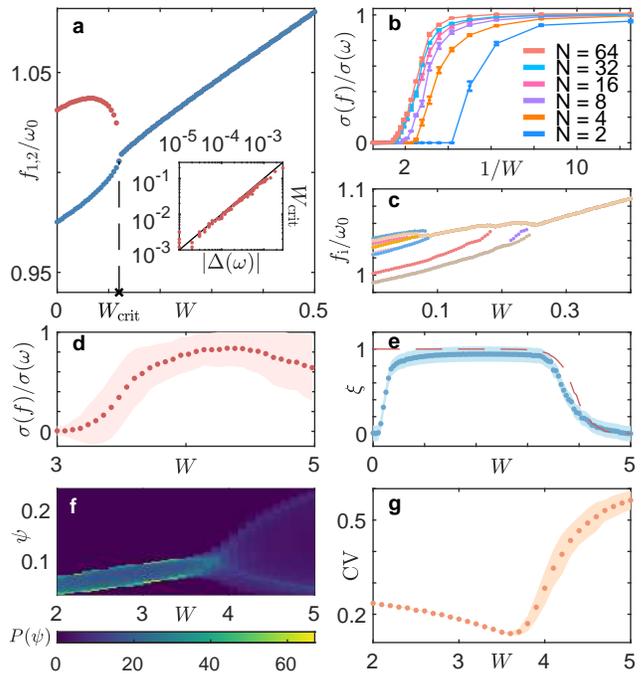}
\caption{
\textbf{The nonlinear nature of interactions between WC oscillators
becomes dominant at stronger coupling, leading to frequency desynchronization in heterogeneous systems,
in addition to loss of coherence.} 
(a) For weak coupling, a pair of interacting oscillators having different intrinsic frequencies $\omega_{1,2}$
exhibit emergent frequencies $f_1, f_2$ (red and blue curves, respectively) which merge at a critical
coupling strength $W_{crit}$. This critical value at which frequency synchronization occurs varies
almost linearly with $\Delta(\omega)$, the difference between the intrinsic frequencies (inset, solid line indicating a slope of $1$).
All emergent frequencies are scaled by the reference frequency $\omega_0$.
(b) The onset of frequency synchronization in globally coupled oscillators (color indicating system size $N$, see legend) 
on increasing the coupling strength $W$ [illustrated explicitly in (c) for $N=10$]. This is indicated by the variation with $1/W$
of the dispersion 
$\sigma(f)$ of the emergent frequencies, normalized by the dispersion $\sigma(\omega)$ of their intrinsic frequencies.
While the spread in the emergent frequencies $f_i$ is comparable to that of the intrinsic frequencies for very weak coupling 
(i.e., $\sigma(f)/\sigma(\omega) \rightarrow 1$ at large $1/W$), $\sigma(f) \rightarrow 0$ on increasing $W$ sufficiently. 
Each curve in (b) is constructed from $10$ realizations. 
(d) Loss of frequency synchronization, indicated by $\sigma(f)/\sigma(\omega)$ becoming finite, occurs at stronger coupling,
as shown for a system of $N=10$ globally coupled oscillators (calculated over $100$ realizations).
In panels (c) and (d), $W$ is increased gradually starting from a random initial state at low $W$.
(e) Onset and subsequent loss of phase synchrony, quantified by the coherence order parameter $\xi$, 
upon increasing $W$ is shown for a system of $10$
heterogeneous (dots) and homogeneous (broken curve) 
oscillators, computed from $500$ realizations. 
(f) The divergence of the synchronization behavior in coupled WC oscillators from that of Kuramoto-like coupled systems at stronger $W$ 
is reflected in the distribution of $\psi$, 
which governs the magnitude of the nonlinear contribution to the coupling (see text). 
The probability densities $P(\psi)$ [see colorbar] are estimated over $500$ realizations.
(g) The coefficient of variation (CV) of $P(\psi)$ increases for $W > 3.6$, suggesting an increased dominance of 
nonlinearity for stronger coupling.
}
\label{fig:fig2}
\end{figure}

In earlier work we have shown that even a system of identical WC oscillators can exhibit a remarkable diversity of collective behavior~\cite{Singh2016,Sreenivasan2017}. 
Apart from ES, which is the only state observed in a homogeneous system of coupled phase oscillators, 
patterns such as quasiperiodic activity (QP),  anti-phase synchronization (APS), and
inhomogeneous in-phase synchronization (IIS) emerge upon increasing the strength of coupling between the WC units.
As shown in Fig.~\ref{fig:fig1}~(e), analogous states are observed on coupling WC oscillators each of which have different intrinsic
frequencies~\cite{noteint}.
Significantly, on increasing $W$, the collective behavior in the case of both heterogeneous and homogeneous systems of globally coupled WC oscillators
is characterized by {\it decreasing} synchrony [as measured by the coherence order parameter $r$, Fig.~\ref{fig:fig1}~(f)].
This is in stark contrast
to the classical result of Kuramoto that
non-identical phase oscillators exhibit coherence at sufficiently strong coupling,
accompanied by the emergence of a common frequency through mutual entrainment~\cite{Kuramoto1984}. 
Fig.~\ref{fig:fig2}~(a) shows that 
the emergent frequencies $f_{1,2}$, resulting from the interaction between a pair of
WC oscillators having distinct intrinsic frequencies $\omega_{1,2}$, converges to a common value at a critical coupling strength $W_{crit}$, 
and then increases with $W$. As seen in the inset,
the minimal coupling strength required for frequency synchronization increases linearly with $\Delta(\omega)$, the extent of variation between the intrinsic frequencies.
The phase transition that marks the onset of synchronization is characterized by using the dispersion of emergent frequencies, $\sigma(f)$ [scaled by
that of the intrinsic frequencies, $\sigma(\omega)$] as an order parameter, and measuring it as a function of $W$.
As seen in Fig.~\ref{fig:fig2}~(b), beginning from very low coupling strengths, increasing $W$ results in all nodes in a system of $N$ coupled WC oscillators eventually converging to a common frequency [$\sigma(f)=0$]
at a critical value that becomes independent of system size at large $N$. 
As shown in Fig.~\ref{fig:fig2}~(c), this comes about through a sequential merging of clusters characterized by a common frequency that are 
formed by synchronization of oscillators that are closest in terms of $\omega$. As $W$ is increased, fusion of clusters that are further apart in their $f$
becomes possible, eventually leading to global frequency synchronization. 
The observation that the global frequency continues to increase with $W$, unlike in a system of coupled phase oscillators,
is characteristic of systems where the periodicity of the globally synchronized state is a function of the
interaction strength between their components. We note that similar phenomena have been observed in 
multiple physiological contexts, e.g., the 
gravid uterus~\cite{Singh2012}.

While the emergence of a common frequency at a finite value of interaction strength in a heterogeneous system of coupled WC oscillators 
may appear similar to phenomena seen in the Kuramoto model, in contrast to the latter system global synchrony is lost in the former
on increasing $W$ further [as indicated by the dispersion of emergent frequencies becoming finite in Fig.~\ref{fig:fig2}~(d)]. 
In addition, 
this transition is accompanied by increased phase desynchronization as indicated by the coherence order parameter 
$\xi =1 - [\langle \sigma_{\max} (\varphi) \rangle /\langle \sigma (\varphi) \rangle ]$ decreasing from its maximum value to zero over the same interval of $W$ [Fig.~\ref{fig:fig2}~(e)]. 
Here, $\sigma$ refers to the dispersion in the phases $\varphi$ of the oscillators, with the maximum dispersion $\sigma_{\max}$
obtained when the WC units are uncoupled (i.e., $W=0$), and $\langle\,\rangle$ indicates time averaging. The instantaneous phase of a WC oscillator is 
defined as $\varphi = \arctan (\{v-\langle v \rangle \}/\{u - \langle u \rangle \})$.

To understand why the behavior of coupled WC oscillators resemble that of Kuramoto-like coupled phase oscillators 
at low coupling but strongly deviate for $W \gtrsim 3$, we investigate how the dynamics of a single unit is affected by interactions 
with neighboring oscillators.
Specifically, for a coupled pair of  WC oscillators $i,j$ , we can express the 
arguments $u_i^{in}, v_i^{in}$ of the nonlinear functions $\mathcal{S}_{u}, \mathcal{S}_{v}$, respectively, in terms of 
a perturbation series expansion of the interaction term $\delta x= W(u_j-v_j)$ around the 
contribution from the local variables, viz., $x=  c_{uu}u_i - c_{uv}v_{i}+ I^{ext}$. When the magnitude of the perturbation relative to the contribution
from the 
local dynamics, $\psi (= \delta x/x)$, is low, the contribution of the nonlinear terms arising from the
interaction will be negligible. 
We note that retaining only the linear term from the series would result in the synchrony between the two oscillators being retained
much beyond the range of $W$ where synchrony is observed when we consider the entire series (see SI). 
As can be seen from Fig.~\ref{fig:fig2}~(f), there is a broadening of the distribution of $\psi$ starting from
$W \approx 3.5$. The accompanying rise of the coefficient of variation (CV) of $\psi$ [Fig.~\ref{fig:fig2}~(g)] implies that
beyond this value of $W$, the
nonlinear terms arising from the interaction with neighboring oscillators begin to dominate.
This suggests that the deviation from synchronization seen around $W = 3.5$ in Fig.~\ref{fig:fig2}~(f) is a consequence of
the nonlinear contribution from the coupling, resulting from the occurrence of large values of $\psi$ with finite probability beyond this point. 
%

\begin{figure}
\includegraphics{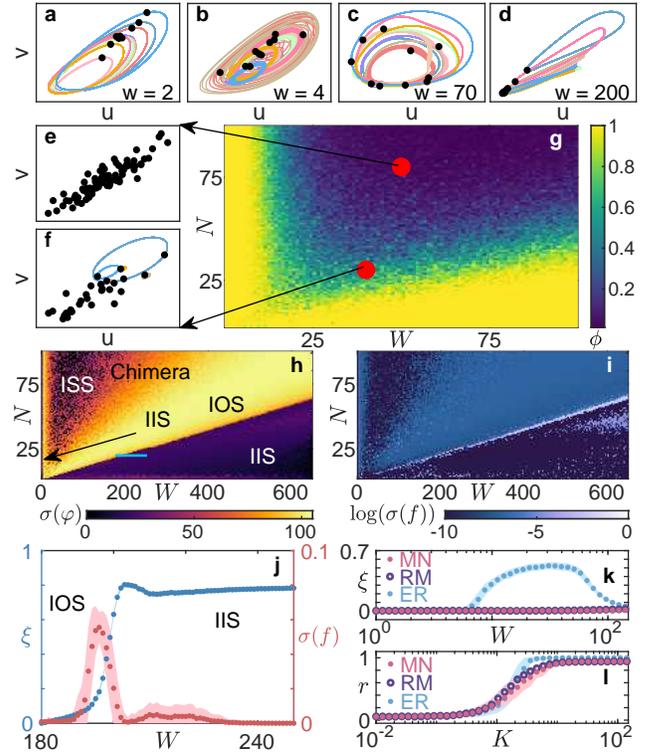}
\caption{{\bf Nonlinearly coupled WC oscillators display diverse collective behavior 
which, in realistic connection topologies, differ significantly from the dynamics of Kuramoto-like coupled systems.}
(a-d) Representative phase-plane portraits for a system of globally
coupled heterogeneous WC oscillators (system size $N=10$) for increasing coupling strength $W$, corresponding 
to inhomogeneous in-phase synchronization (IIS,
a: low $W$; d: high $W$),
inhomogeneous quasiperiodic (IQP, b) and inhomogeneous out-of-phase synchronization
(IOS, c) states. (e-f) For 
larger systems, inhomogeneous steady states (ISS, e), and chimera states which comprise
coexisting oscillating and static elements (f), can also
be observed. 
Filled circles indicate individual oscillators.
(g) Varying $N$ and/or $W$ can arrest the activity of a finite number of oscillators
(ISS being the limiting case), displayed in terms of the fraction of active oscillators $\phi$
(averaged over $10$ realizations), with chimera states seen for $0 < \phi < 1$.
(h-i) On increasing $W$ further, the variations observed in phase and frequency synchronization
are indicated by the corresponding dispersions
(h: $\sigma(\varphi)$; i: $\sigma(f)$)~\cite{note2}.
(j) Onset of phase synchrony, measured by the coherence order parameter $\xi$, during the passage from IOS to IIS (high $W$) regimes, 
is 
accompanied by loss of frequency synchronization in the transition region [marked by a horizontal bar in (h)], as indicated by the peak in $\sigma(f)$.
Results shown for $N=20$ and averaged over $500$ realizations.
(k-l) The impact of the nature of coupling on synchronization in
sparse networks of heterogeneous oscillators is examined for
an empirical network (MN) representing the Macaque connectome ($N=266$)~\cite{Pathak2020}, its degree-preserved randomized surrogate (RM) and 
an Erd\H{o}s-R\'{e}nyi random network (ER) of same size and connection density.
(k) For the nonlinearly coupled system of WC oscillators, phase coherence, measured by $\xi$, is observed only in ER 
networks at an intermediate range of $W$. (l) In a system of coupled phase oscillators, coherence ($r \sim 1$) is seen in all three networks for sufficiently high $W$.  
Results in (k-l) are averaged over $200$ realizations.
}
\label{fig:fig3}
\end{figure}

For a globally coupled system of $N$ heterogeneous WC oscillators we observe collective
dynamical transitions analogous to those described above for a coupled pair on increasing $W$ [see Fig.~\ref{fig:fig3}~(a-d)]~\cite{note1}.
In addition, for larger $N$ we observe patterns in which the coupling arrests the activity of a finite number of oscillators [Fig.~\ref{fig:fig3}~(e-f)].
Defining $\phi$ as the fraction of
oscillating nodes, shown as a function of $N$ and $W$ in Fig.~\ref{fig:fig3}~(g), we can classify these into time-invariant Inhomogeneous Steady States (ISS)
[$\phi=0$, see panel (e)] and ``Chimera'' patterns characterized by coexistence
of oscillating and non-oscillating units [$0<\phi<1$, see panel (f)]. 
While phase and frequency synchronization regimes largely seem to overlap [compare Fig.~\ref{fig:fig3}~(h) and (i)], panel (j) shows a striking exception
seen during the transition from IOS to IIS regimes. As coupling becomes stronger, the increase in phase synchrony ($\xi$) is accompanied by
a transient loss in frequency synchronization indicated by the peak in $\sigma(f)$.
 

Having established that 
increasing the strength of nonlinear interactions 
between globally coupled neural oscillators may result in a transition to counter-intuitive patterns of collective dynamics,
viz., {\it loss} of frequency synchronization and phase coherence, we now examine the generality of this result for 
different network topologies. 
As our comparative study of such systems with Kuramoto-like coupled phase oscillators is motivated by the phenomenon
of long-range synchronization in brain activity,
we specifically consider a network representing 
the Macaque connectome (MN) comprising $N=266$ brain areas~\cite{Pathak2020}.
We place WC oscillators on each node, whose directed connections  are distributed exponentially with mean degree $\langle k^{MN} \rangle=9.78$,
and compare its collective dynamics with that obtained from randomized surrogate networks (RM) having the same degree sequence 
as MN, and Erd\H{o}s-R\'{e}nyi random networks (ER) having the same size and mean degree~\cite{SuppInfo}.
As shown in Fig.~\ref{fig:fig3}~(k), the empirical network always exhibits
phase desynchronization ($\xi \approx 0$), as do the RM networks,
while partial synchronization is seen in ER networks. 
In contrast, for coupled phase oscillators [see Fig.~\ref{fig:fig3}~(l)] we observe a continuous transition to complete phase synchronization
($r=1$) at sufficiently high $W$, independent of the network topology.

The distinct collective dynamics manifested by neural oscillators placed at the vertices of the empirical network,
as compared to networks having different connection topologies,
assume importance in view
of the physiological implications of such behavior. The nonlinearity-driven transition to decoherence could point towards an explanation 
of reports linking the loss of consciousness
with increased synchrony of activity, e.g., as seen during epilepsy~\cite{Arthuis2009}. Indeed, these phenomena can be viewed
as outcomes of the decreased interaction between brain areas, as suggested by our results described here.
In contrast to Kuramoto-like coupled phase oscillators which show complete synchronization regardless of the connection topology, 
the collective dynamics of systems of neural oscillators underline the role that network structure plays in shaping the emergent activity
of the brain.
Our results, highlighting the
importance of nonlinear interactions that manifest at stronger coupling strengths, is an attempt at building
a more appropriate paradigm for describing strongly nonlinear complex adaptive systems
by establishing a phenomenology of the associated dynamical transitions. 

\begin{acknowledgements}
\small{
We would like to thank Shivakumar Jolad, Chandrashekar Kuyyamudi, Anand Pathak and Amit Reza for helpful discussions.
SNM has been supported by the IMSc Complex Systems Project (12th 
Plan), and the Center of Excellence in Complex Systems and Data 
Science, both funded by the Department of Atomic Energy, Government of 
India. The simulations required for this work were 
supported by IMSc High Performance Computing facility (hpc.imsc.res.in) [Nandadevi].}
\end{acknowledgements}

\clearpage
\FloatBarrier

\onecolumngrid

\setcounter{figure}{0}
\renewcommand\thefigure{S\arabic{figure}}  
\renewcommand\thetable{S\arabic{table}}

\vspace{1cm}

\begin{center}
\textbf{\large{SUPPLEMENTARY INFORMATION}}\\

\vspace{0.5cm}
\textbf{\large{The nonlinearity of interactions drives networks of neural oscillators to decoherence at strong coupling}}

\vspace{0.5cm}
\textbf{Richa Tripathi, Shakti N. Menon and Sitabhra Sinha}
\end{center}

\section*{List of Supplementary Figures}

\begin{enumerate}
\item Fig S1: Dependence of the intrinsic frequency $\omega$ of a WC oscillator on parameters controling the nodal dynamics.
\item Fig S2: Absence of refractoriness results in only a marginal decrease in coherence underlining the role of nonlinearity
in driving the transition to desynchronization at stronger coupling in a system of WC oscillators.
\item Fig S3: The strength $W$ of coupling between two WC oscillators alters their collective dynamics, showing that the loss of
coherence coincides with the increasing dominance of nonlinear terms in the coupling.
\item Fig S4:  The distributions of in-, out- and total degree of nodes in the macaque brain network follows an exponential nature,
quite distinct from that of Erd\H{o}s-R\'{e}nyi random networks having the same size and average degree.
\item Fig S5: For a network of $N (=266)$ coupled phase oscillators, the transition to coherence upon increasing the coupling strength 
$K$ is qualitatively similar for different connection topologies.
\item Fig S6: The phase coherence and frequency synchronization behaviors of a network of $N (=266)$ nonlinearly coupled WC oscillators are dependent on the connection topology.
\end{enumerate}

\section*{Introducing heterogeneity in the intrinsic oscillatory dynamics of the nodes}

\begin{figure}
\includegraphics{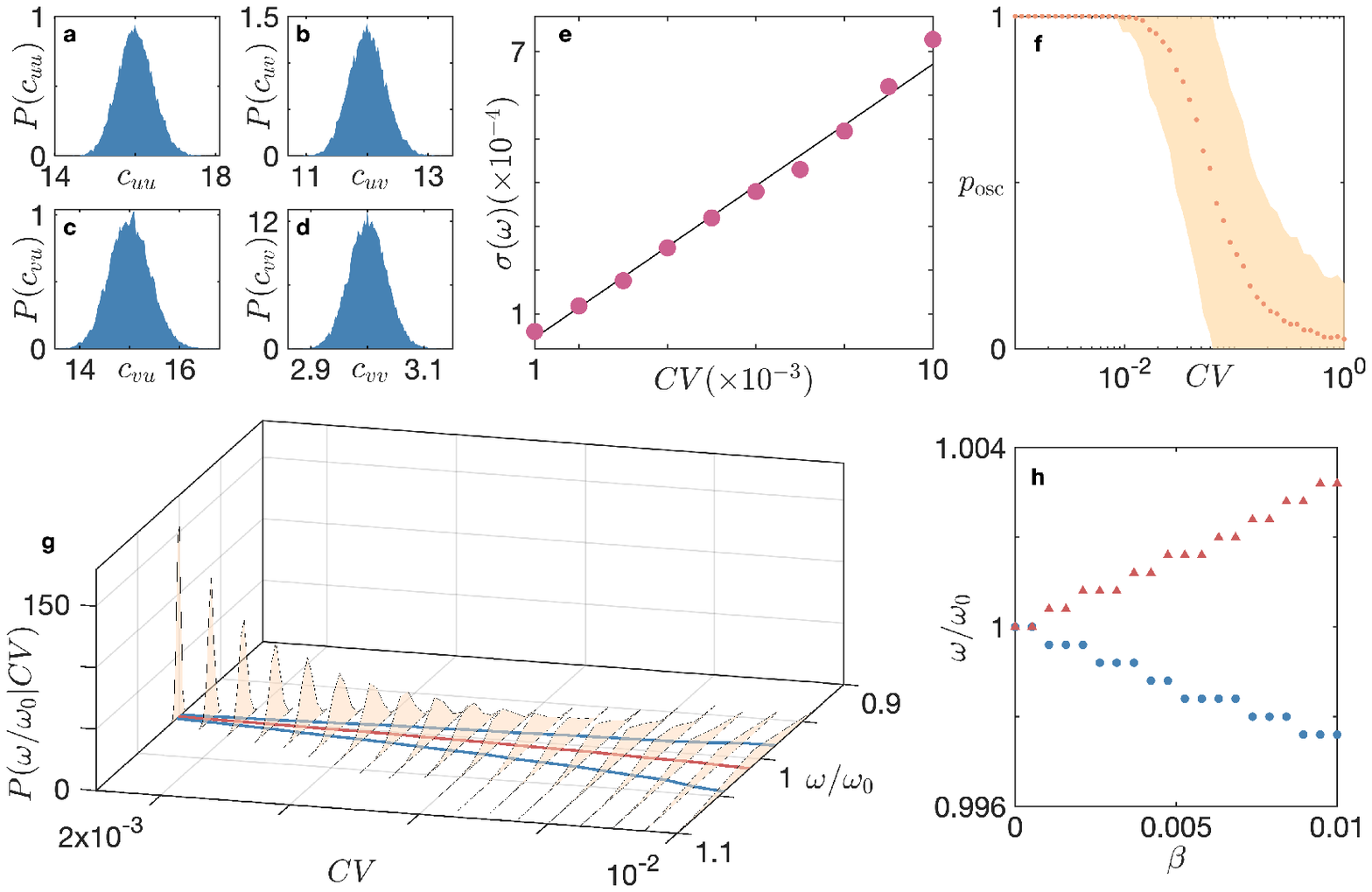}
\caption{
{\bf Dependence of the intrinsic frequency $\omega$ of a WC oscillator on parameters controling the nodal dynamics.}
(a-d)~The parameters $c_{\mu\nu}$ ($\mu, \nu \in \{u,v\}$) governing the interactions within and between the excitatory ($u$) and inhibitory ($v$) subpopulations are sampled from log-normal distributions having coefficient of
variation $CV=0.01$ and mean value (a)~$\langle c_{uu} \rangle=c^*_{uu} =16$, (b)~$\langle c_{uv} \rangle =  c^*_{uv} =12$,
(c)~$\langle c_{vu} \rangle =  c^*_{vu} = 15$ and (d)~$\langle c_{vv} \rangle= c^*_{vv} = 3$. The oscillator obtained upon choosing the mean values for
all the parameters has a frequency $\omega_0$ used as the reference frequency relative to which all frequencies are expressed.
The probability distributions shown are generated from an ensemble of $2\times 10^{4}$
realizations. (e)~The dispersion of $\omega$ for the ensemble of oscillators, constructed by sampling the parameters
$c_{\mu\nu}$ ($\mu, \nu \in \{u,v\}$) from a distribution, increases almost linearly with the $CV$. Each data point is calculated from
$2000$ realizations. The linear fit (shown using a solid line) corresponds to a slope of $0.07$. 
(f) A WC node whose parameters are chosen from the distributions shown in (a-d) is almost certain to exhibit oscillations
when the CV of the distributions is small. However, the probability $p_{\rm osc}$ that it will oscillate decreases sharply beyond $CV=10^{-2}$ (for which $p_{\rm osc} \sim 0.98$), becoming negligibly small by $CV=1$.
The mean and dispersion of $p_{\rm osc}$ for each value of CV are estimated from $10^3$ realizations.
(g) The probability distribution P($\omega/\omega_0|CV)$ of the intrinsic frequencies scaled by the reference frequency, given the CV 
of the distributions from which the WC node parameters are sampled, constructed from $5000$ realizations for each value of CV and smoothened
using a Gaussian kernel.
The red and blue curves at the base show the variation of the mean value and standard deviation (respectively) of $\omega/\omega_0$ as a 
function of CV.
(h) In spite of the complex interactions between the subpopulations within a WC oscillator, its scaled intrinsic frequency $\omega/\omega_0$ exhibits
an unexpectedly simple linear relation with $\beta$, the relative difference between the parameters $c_{\mu\nu}$  from their
reference values $c^*_{\mu\nu}$, with the cases $c_{\mu\nu} < c^*_{\mu\nu}$ and $c_{\mu\nu} > c^*_{\mu\nu}$ represented by
red triangles and blue circles, respectively. 
}
\label{fig:figS1}
\end{figure}

In the main text, we have reported the collective dynamics of systems of coupled phase oscillators, as well as, that of
Wilson-Cowan (WC) oscillators which represent the interactions between excitatory and inhibitory pools of neurons in a
brain region~\cite{Wilson1972}. In the classical treatment of coherence in phase oscillators pioneered by Kuramoto~\cite{Kuramoto1984}, an essential ingredient
is the heterogeneity of the oscillators. This is usually implemented by the intrinsic frequencies of the oscillators 
being randomly sampled from a distribution.     
Therefore, to allow for comparison between the results obtained from the systems of coupled phase oscillators and that of the nonlinearly coupled
WC oscillators, we have introduced heterogeneity in the intrinsic oscillation frequencies $\omega$ of the latter.
This is implemented by varying the internal
coupling parameters $c_{\mu \nu}$ $(\mu, \nu = u, v)$ in Eqn.~(1). 
The values of each of these four parameters are independently chosen from log-normal distributions having a specified 
mean ($\langle c_{\mu \nu} \rangle$) and 
coefficient of variation ($CV$) [Fig.~\ref{fig:figS1}~(a)-(d)]. For all simulations reported in the main text, we have chosen
$\langle c_{uu} \rangle =16$,  $\langle c_{uv}  \rangle=12$, $\langle c_{vu}  \rangle=15$ \& $\langle c_{vv}  \rangle=3$,
and $CV = 0.01$.  
The frequencies $\omega$ of each of the oscillators can be compared with the reference frequency $\omega_0$ obtained for the oscillator
having  $c_{\mu \nu} = \langle c_{\mu \nu} \rangle \forall \mu,\nu$.
Fig.~\ref{fig:figS1}~(e) shows that the dispersion of these frequencies, $\sigma(\omega)$,
scales linearly with the $CV$ of the distributions. 
Beyond $CV=0.01$, depending on the choice of the parameters
$c_{\mu \nu}$ $(\mu, \nu = u, v)$, a WC unit may not exhibit oscillations, i.e., the oscillation probability $p_{\rm osc} < 1$.
Hence, the choice of $CV = 0.01$ ensures that all nodes in an ensemble are intrinsically capable of oscillations while also maximizing the
extent of heterogeneity [Fig.~\ref{fig:figS1}(f)]. 
Furthermore, for a given choice of $CV$, the 
frequencies (scaled by the reference frequency $\omega_0$)
follows a log-normal distribution [Fig.~\ref{fig:figS1}~(g)]. 
To further elucidate the role of
$c_{\mu \nu}$ in determining the intrinsic oscillator frequencies, we have also performed simulations
wherein the parameters are chosen deterministically to have the values $\langle c_{\mu \nu} \rangle (1\pm CV)$. For the case
where the parameters are above (below) the corresponding
mean values, we observe that the intrinsic frequency $\omega$ of the WC oscillator decreases (increases) as
$CV$ is increased.
[Fig.~\ref{fig:figS1}~(h)].

\section*{The role of nonlinear interactions on the collective dynamics}
\begin{figure}
\includegraphics{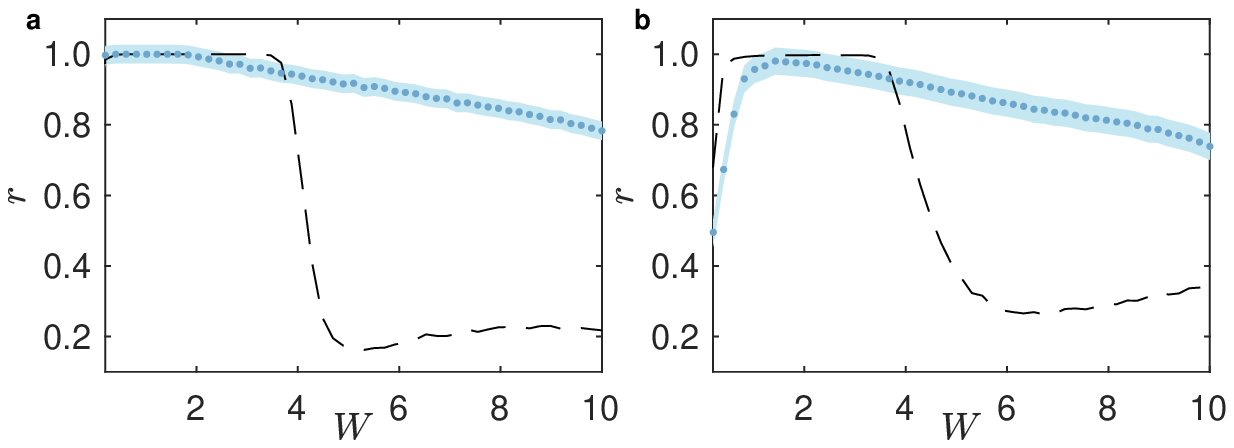}
\caption{{\bf Absence of refractoriness results in only a marginal decrease in coherence underlining the role of nonlinearity
in driving the transition to desynchronization at stronger coupling in a system of WC oscillators.} 
Results are displayed for a system of $N=10$ globally coupled WC oscillators where the oscillators could either (a) have 
identical intrinsic frequency $\omega$ (homogeneous) or (b) are heterogeneous.
The blue dots and shaded region
represent the means and standard deviations of the coherence order parameter $r$ for simulations done with the refractory
periods $r_u,r_v =0$ computed over $200$ realizations, while the broken curves correspond to the results displayed in Fig.~1~(f) in the main text (for which $r_u,r_v =1$).
}
\label{S8}
\end{figure}
\subsection*{Finite refractory period}
In the main text we have mentioned that choosing finite values for the refractory periods of the neurons makes the interactions
between the WC oscillators irreducibly nonlinear. If the refractoriness of the components is neglected by choosing $r_u, r_v = 0$,
 it implies that the entire complement of neurons belonging to the excitatory and inhibitory subpopulations in each 
node is available for activation at each instant. Thus, the interaction - represented by the second term of the evolution equations
[Eqn.~(1)] - becomes (upto multiplication by a constant factor) simply
a sigmoid function of a linear combination of the variables describing the system. By suitable choice of parameters one can
operate in the linear range of the sigmoid function, thereby rendering the model effectively linear. Fig.~\ref{S8} shows that
the behavior of the globally coupled WC oscillator system is indeed remarkably different when the refractoriness is neglected
compared to the case when a finite refractory period is considered. In particular, the sharp transition to decoherence when the
coupling $W$ is increased is no longer observed, regardless of whether we are considering a homogeneous or heterogeneous 
system of oscillators. This phenomenon underlines the critical importance of nonlinearity arising from the refractory property
of neurons in driving the transition to decoherence at strong coupling.
 
\subsection*{Nonlinear contributions from the sigmoid interaction function}
In the main text we have mentioned that the divergence at stronger coupling of the behavior of coupled WC oscillators from that of sinusoidally
coupled phase oscillators can be seen as an outcome of the increased contribution of nonlinear terms arising from the interaction
between oscillators. To this end, we perform a power series expansion of the sigmoid interaction functions $\mathcal{S}_u, \mathcal{S}_v$ in
Eqn.~(\ref{wc_eqn}) and obtained a reduced model by retaining only the linear terms in the expansion.

We can illustrate this procedure for a pair of coupled WC oscillators, where the time-evolution of the variable describing the behavior
of the excitatory sub-population of first WC unit is described by
$$\tau_u \dot{u}_1 = -u_1 + (\kappa_v - r_u u_1) {\cal S}_u(c_{uu}u_{1} - c_{vu}{v}_{1} + W(u_{2} - v_{2}) + I_{u_1}),$$
or, equivalently,
$$\tau_u \dot{u}_1 = -u_1 + (\kappa_v - r_u u_1) {\cal S}_u(X_{0} + \Delta X),$$
where $\Delta X = W(u_{2} - v_{2})$ is the perturbation around $X_{0} = c_{uu}u_{1} - c_{vu}{v}_{1} + I_{u_1}$.
Linearization around $X_0$ yields
$$ {\cal S}_u(X_{0} + \Delta X) = {\cal S}_u(X_0) + {\cal S}_u'(X)|_{X = {X_0}}\Delta X.$$
Similar linearized forms can also be obtained for $u_2$, $v_1$ and $v_2$.
The numerical solution of these equations with only the linear contribution in the interaction term yields the dynamics of the
reduced model. By comparing these results with that of the WC model [Eqn.~(\ref{wc_eqn})], the role played by the nonlinear terms at stronger 
coupling can be made explicit.
 
\begin{figure}
\includegraphics{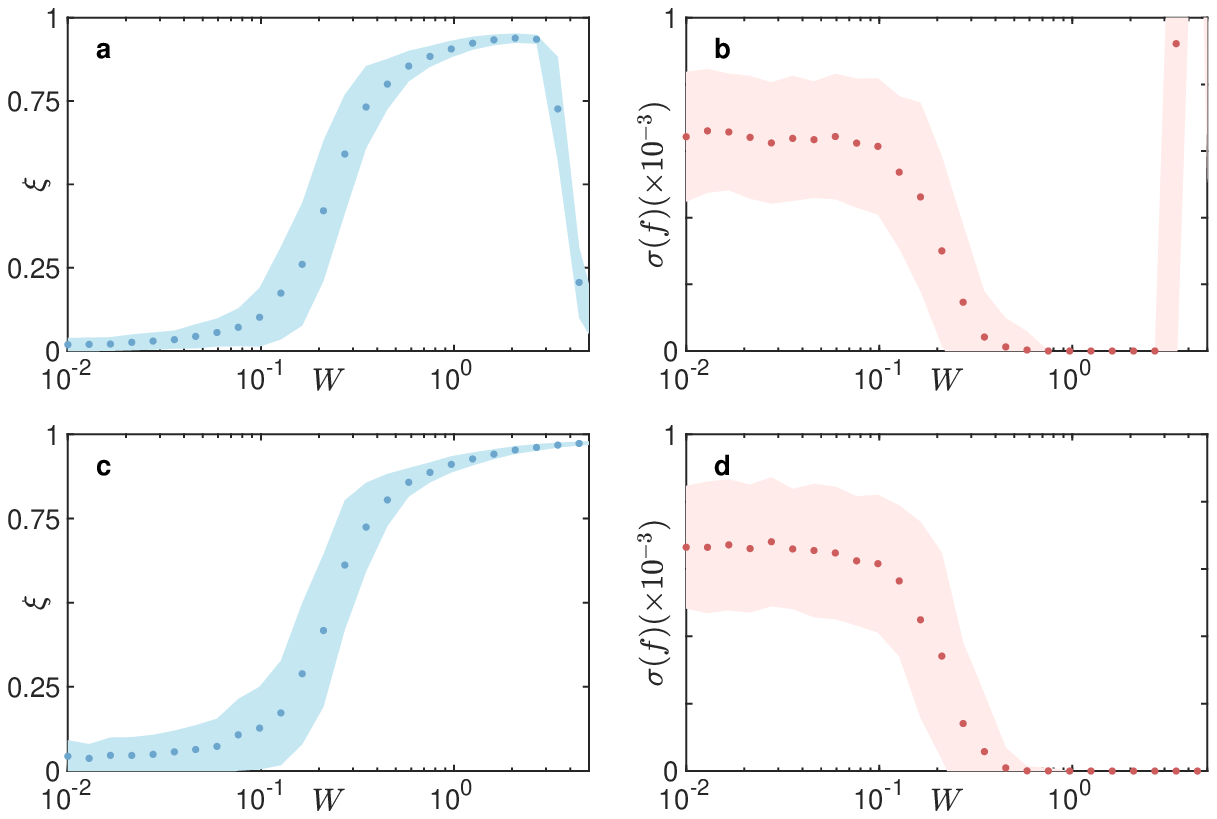}
\caption{{\bf Decoherence and loss of frequency synchronization observed at stronger coupling in a heterogeneous system of globally coupled WC oscillators is
a consequence of increased contribution from nonlinear interaction terms.} We show the variation with $W$ of (left) the coherence order parameter $\xi$ 
and (right) the dispersion $\sigma(f)$ of the emergent frequencies in a fully connected network of $N=10$ oscillators described by (top row) the WC
model [Eq.~(1) in the main text] and (bottom row) a reduced model obtained by retaining only the linear part of the perturbation series expansion of the interaction term.
It can be seen that increasing the coupling strength beyond $W\sim 3$ results in the WC model (which contains nonlinear interaction terms) losing both 
phase coherence and frequency synchrony, while the network with the reduced, linearized model still remains coherent and synchronized. The
dots and the shaded regions represent the mean and standard deviation of the observables calculated over $500$ realizations at each value of $W$.}
\label{S3}
\end{figure}

Figure~\ref{S3} shows such a comparison of the coherence and frequency synchronization behavior of a globally coupled network 
of $N = 10$ WC units (top panel) with an equivalent reduced model network having linearized interactions between the units. While the two models
show qualitatively identical responses to increased coupling when $W$ is low, at higher values of $W (\sim 3)$ they diverge. While the WC model
loses both coherence and frequency synchrony, this is not seen to be the case for the reduced model. As the two models differ only in terms
of the absence of nonlinear interaction terms in the reduced model, we can conclude that the distinct behavior of the WC model (in comparison
to the linearized reduced model, as well as, Kuramoto-like coupled phase oscillators) owes its
origin to the increased importance of nonlinear interactions terms at this higher value of $W$.


\FloatBarrier




\section*{Collective dynamics in networks having different connection topologies}

\begin{figure}
\includegraphics{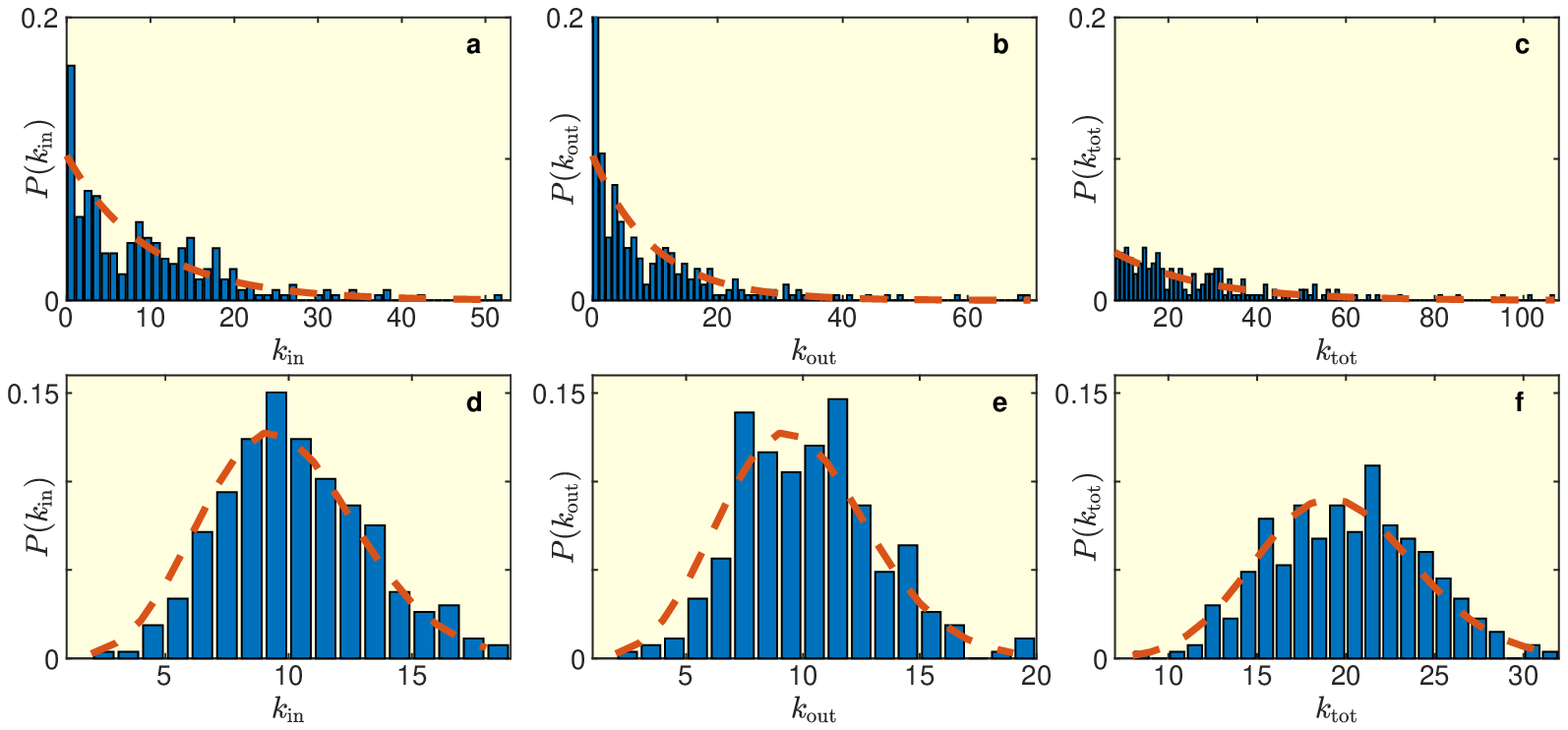}
\caption{{\bf The distributions of in-, out- and total degree of nodes in the Macaque brain network (top row) follows an exponential nature,
quite distinct from that of Erd\H{o}s-R\'{e}nyi random networks having the same size and average degree (bottom row).}
The probability distributions of the incoming ($k_{in}$: a, d), outgoing ($k_{out}$: b, e) and aggregate ($k_{tot}$: c, f) number of connections for the nodes are shown along
with the maximum likelihood estimates of the best fit exponential (top row) and Poisson (bottom row) distributions which are shown as broken curves.}
\label{S6}
\end{figure}

\noindent
{\bf Statistics of probability distribution of degrees for Macaque brain network and ER surrogate network.}\\
We have shown in Fig.~\ref{S6} (top row) the distributions for the in-coming, out-going and total number of connections 
for the nodes in the network derived from the Macaque connectome. These empirical distributions appear to be exponential in nature, as indicated by the fitted curves
(obtained by maximum likelihood estimation). For comparison, we show (bottom row) the corresponding distributions for 
random networks having the same size $N(=266)$ and 
average degrees $\langle k_{in} \rangle (= 9.78)$, $\langle k_{out} \rangle (= 9.78)$ and $\langle k_{tot} \rangle (= 19.56)$ as the empirical network.\\

\noindent
{\bf Construction of the surrogate networks for the macaque brain network.}\\
To investigate the role of network topology in the collective behavior (specifically, coherence and frequency synchronization) 
of a system of WC oscillators arranged on the network derived from the Macaque connectome, we have compared the observations
from the empirical network with four surrogate ensembles (each comprising $500$ network realizations, with network size $N=266$) of the following types:
\begin{enumerate}
\item Degree-Preserved Module-Preserved randomized macaque brain network (DPMP): obtained by performing degree-preserved randomizations within each module of the macaque brain network. In other words, directed edges between two randomly chosen distinct pairs of nodes are swapped such that 
the degree and module membership of each of the nodes are preserved. For each realization of a surrogate network we perform 
$5*E$ (E = number of edges in the macaque brain network) edge swaps. 
\item Degree-Preserved randomized macaque brain network (DP): obtained by performing degree-preserved randomization as above, 
but without considering the module membership of the pairs of nodes that were chosen for an edge swap operation. Thus,
in these surrogate networks, only the degree sequence of the corresponding nodes in the Macaque brain network has been preserved. 
For each realization of a surrogate network of this type we perform $2*E$ edge swaps.
\item Module-Size-Preserved random network (MSP): obtained by generating modular networks having the same average degrees 
($\langle k_{in} \rangle$, $\langle k_{out} \rangle$ and $\langle k_{tot} \rangle$) and number of modules ($5$) as the
macaque brain network, with each module having the same size (viz., $54,\ 71,\ 60,\ 39$ and $42$) as the corresponding module in the empirical network.        \item Erd\H{o}s-R\'{e}nyi random networks (ER): obtained by generating homogeneous random networks with the uniform probability of
connection between any pair of nodes set equal to that of the mean connection probability in the macaque brain network (ensuring that
the average degrees $\langle k_{in} \rangle$, $\langle k_{out} \rangle$ and $\langle k_{tot} \rangle$ are preserved).
\end{enumerate}

\begin{figure}
\includegraphics{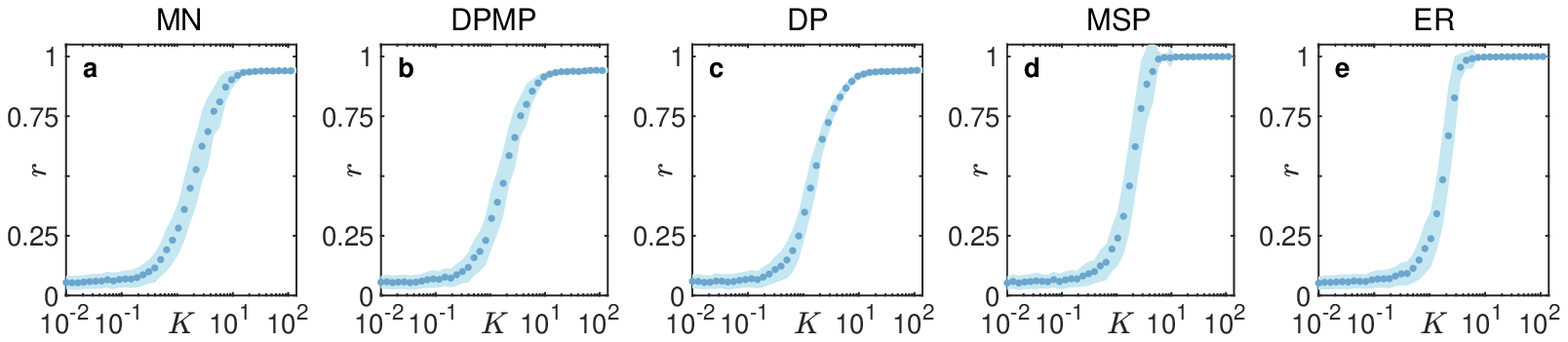}
\caption{{\bf For a network of $N (=266)$ coupled phase oscillators, the transition to coherence upon increasing the coupling strength 
$K$ is qualitatively similar for different connection topologies.} The coherence order parameter $r$ is shown as a function of $K$ for 
 (a) the Macaque brain network (MN), (b) randomized surrogates of MN that preserves degree sequence and module membership of the
nodes (DPMP), (c) randomized surrogates of MN preserving only the degree sequence (DP), (d) random modular networks whose module sizes 
and average degree are same as that of MN (MSP), and (e) Erd\H{o}s-R\'{e}nyi random networks having average degree same as that of MN (ER).
In each case, the mean (represented by the filled circles) and standard deviation (indicated by the shaded interval) is calculated over an ensemble of $500$ realizations.}
\label{S4}
\end{figure}

\begin{figure}
\includegraphics{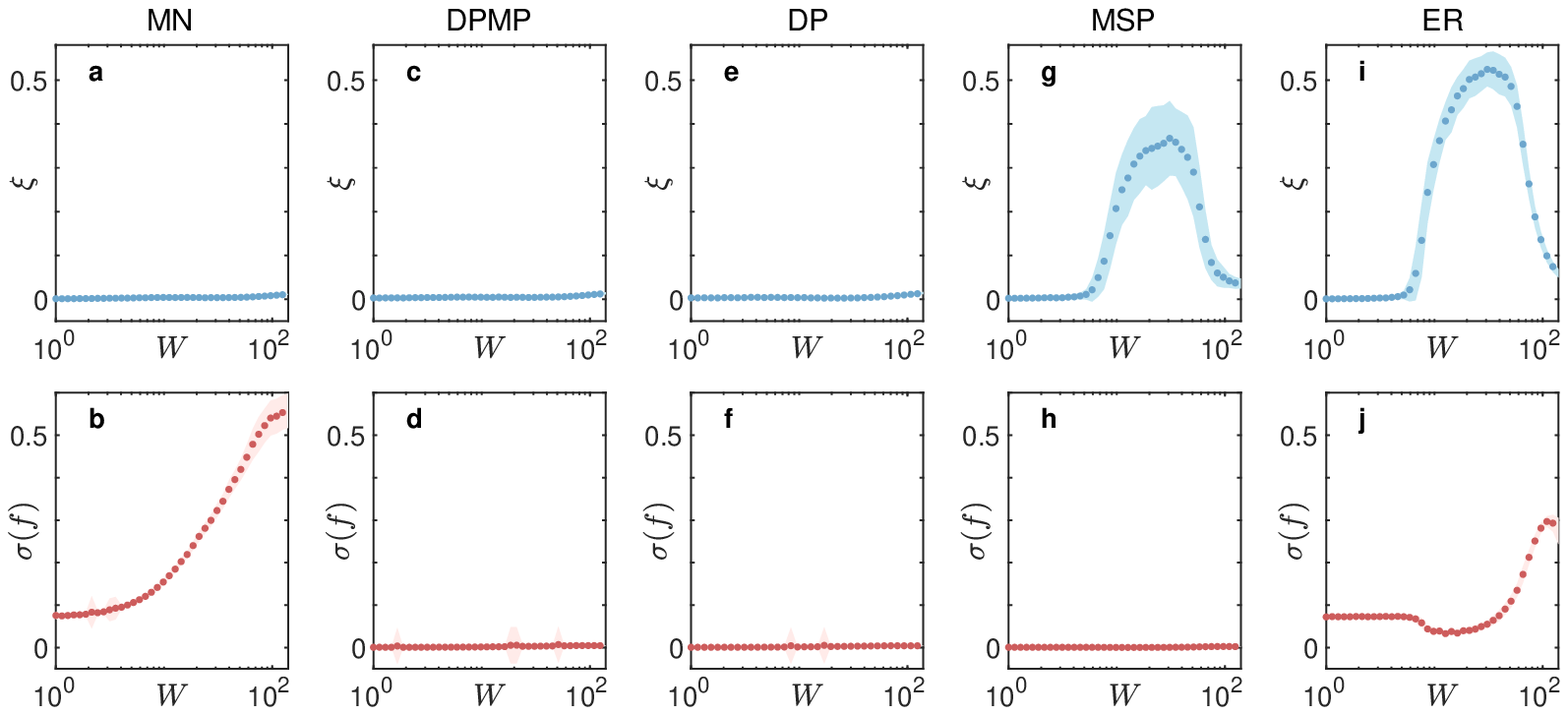}
\caption{{\bf The phase coherence (top row) and frequency synchronization (bottom row) behaviors of a network of $N (=266)$ nonlinearly coupled WC oscillators are dependent on the connection topology.} The coherence order parameter $\xi$ and the dispersion of emergent frequencies $\sigma(f)$
are shown as a function of the interaction strength $W$ for  (a-b) the Macaque brain network (MN), (c-d) randomized surrogates of MN that preserves degree sequence and module membership of the
nodes (DPMP), (e-f) randomized surrogates of MN preserving only the degree sequence (DP), (g-h) random modular networks whose module sizes 
and average degree are same as that of MN (MSP), and (i-j) Erd\H{o}s-R\'{e}nyi random networks having average degree same as that of MN (ER).
In each case, the mean (represented by the filled circles) and standard deviation (indicated by the shaded interval) is calculated over an ensemble of $500$ realizations. Note that there is complete absence of coherence at sufficiently large values of $W$ for all the connection topologies.}
\label{S5}
\end{figure}

In Fig.~\ref{S4} we show the variation of coherence (measured by the order parameter $r$) with the coupling strength $K$ for the macaque
brain network and the four ensembles of surrogate networks, when phase oscillators are placed at each node and sinusoidally coupled to
their network neighbors. Fig.~\ref{S5} shows both coherence (measured by $\xi$, see top row) and frequency synchronization (measured 
by $\sigma(f)$, see bottom row) in the corresponding networks, when nonlinearly coupled WC oscillators are placed on the nodes.
\\



\begin{thebibliography}{}
\bibitem{note1}
While for small $N$, frequency and phase synchrony can be regained upon increasing the coupling further, as system size increases
the value of $W$ at which this occurs becomes extremely high.
\bibitem{note2}
Note that in Fig.~\ref{fig:fig3}~(h-i), the standard deviations
were taken over the oscillating nodes only in the ``Chimera'' region, and in the ISS region the values were set as $\sigma (\varphi) = 0$ and $\log(\sigma (f)) = -10$, respectively.
\end{thebibliography}
\end{document}